\documentclass[letterpaper, 10 pt, conference]{ieeeconf}  

\IEEEoverridecommandlockouts                              
\usepackage{syntax}
\usepackage{url}
\usepackage{graphicx}
\usepackage{amssymb}

\usepackage{amsfonts}
\usepackage[english]{babel}
\usepackage[bottom]{footmisc}
\usepackage{algorithm}
\usepackage{algorithmicx}
\usepackage{algpseudocode}
\usepackage{amsthm}
\usepackage{siunitx}
\usepackage{array}
\usepackage{verbatim}
\usepackage{subcaption}
\usepackage{listings}
\usepackage{amsmath}
\usepackage{adjustbox}
\usepackage{csquotes}
\usepackage{microtype}
\usepackage{graphicx}
\usepackage{amsmath}
\usepackage{todonotes}
\usepackage{xcolor}
\usepackage{cite}
\usepackage{amsfonts}
\usepackage{xspace} 
\usepackage{multicol}
\graphicspath{{images/}}

\newtheorem{theorem}{Theorem}

\newtheorem{problem}{Problem}

\newtheorem{assumption}{Assumption}
\newtheorem{proposition}{Proposition}

\newtheorem{definition}{Definition}

\title{\LARGE \bf Synergistic Offline-Online Control Synthesis \\via Local Gaussian Process Regression}

\author{John Jackson$^1$, Luca Laurenti$^2$, Eric Frew$^1$, and Morteza Lahijanian$^1$
\thanks{This work was supported in part by the NSF grant 2039062 and NSF Center for Unmanned Aircraft Systems under award IIP-1650468.}%
\thanks{$^1$ Smead Aerospace Engineering Sciences,
        University of Colorado Boulder, CO, USA.
        {\tt \small \{john.m.jackson, eric.frew, morteza.lahijanian\}@colorado.edu}
        }%
\thanks{$^2$ Center of Systems and Control, TU Delft University, The Netherlands.
        {\tt \small l.laurenti@tudelft.nl}}%
}

\newcommand{\reals}{\mathbb{R}}

\newcommand{\naturals}{\mathbb{N}}

\DeclareMathOperator*{\argmax}{arg\,max}
\newcommand\indicator{\mathbf{1}}
\newcommand{\interindicator}[2]{\indicator^{#2}_{#1}}

\newcommand{\x}{\textbf{{x}}}   

\renewcommand{\u}{\textnormal{\textbf{u}}}
\newcommand{\policy}{\pi}

\newcommand{\procnoise}{\textnormal{\textbf{w}}}

\newcommand{\dataset}{\mathrm{D}}
\newcommand{\distribution}{p}

\newcommand\GPFull{\mathcal{GP}}
\newcommand\GPLocal{\GPFull_{x_k}}
\newcommand\GPFullSet{\GPFull^{\Apimdp}}
\newcommand\GPLocalSet{\GPFullSet_{x_k}}
\newcommand{\nns}{\ell}

\newcommand{\APs}{AP}
\newcommand\spec{\varphi}

\newcommand{\IMDPText}{\textsc{IMDP}\xspace}
\newcommand{\IMDP}{\mathcal{I}}
\newcommand{\PIMDPText}{\textsc{PIMDP}\xspace}
\newcommand{\PIMDP}{\mathcal{P}}
\newcommand{\Qimdp}{Q}
\newcommand{\qimdp}{q}
\newcommand{\qpimdp}{s}
\newcommand{\Qpimdp}{S}
\newcommand{\Pup}{\hat{P}}
\newcommand{\Plow}{\check{P}}
\newcommand{\Aimdp}{A}
\newcommand{\aimdp}{u}
\newcommand{\Apimdp}{A}
\newcommand{\apimdp}{u}

\newcommand{\PossStates}{\Qpimdp^a(\qpimdp')}

\newcommand{\adversary}{\xi}
\newcommand{\Adversary}{\Xi}

\newcommand{\DFAText}{\textsc{DFA}\xspace}
\newcommand{\DFAA}{\mathcal{A}_\spec}
\newcommand{\DFAStates}{Z}

\newcommand{\DFATransition}{\delta}
\newcommand{\DFAState}{z}
\newcommand{\DFAFinalState}{\DFAStates_F}

\newcommand{\policyoff}{\policy_{\text{off}}}

\newcommand{\plow}{\underline{p}}
\newcommand{\pupp}{\overline{p}}

\renewcommand{\u}{\textnormal{\textbf{u}}}

\newcommand{\optpolicy}{\policy^*}

\newcommand{\Setofpolicies}{\Pi}
\newcommand{\FullSet}{X}

\newcommand{\distBound}{\epsilon}

\newcommand{\gest}{\hat f}
\renewcommand{\path}{\omega}
\newcommand{\pathX}{\path_{\x}}

\newcommand{\PathFin}{\Omega_{\x}^{\text{fin}}}
\newcommand{\kernel}{k}
\newcommand{\ROISet}{R}
\newcommand{\roi}{r}

\newcommand{\Spec}{\varphi}
\newcommand{\LTLf}{LTLf\xspace}
\newcommand{\LTL}{LTL\xspace}
\newcommand{\Globally}{\mathcal{G}}
\newcommand{\Eventually}{\mathcal{F}}

\newcommand{\Next}{\mathcal{X}}
\newcommand{\trace}{\rho}

\newcommand{\Prop}{\mathfrak{p}}

\renewcommand{\kernel}{\kappa}
\newcommand{\mean}{\mu}

\newcommand{\imdp}{\textsc{IMDP}\xspace}

\newcommand{\PIMDPS}{\mathcal{P}}

\newcommand{\I}{\mathcal{I}}

\newcommand{\probDist}{\mathcal{D}}

\newcommand{\feasibleDist}[2]{\gamma_{#1}^{#2}}

\newcommand{\Amdp}{A}

\newcommand{\amdp}{a}
\newcommand{\qmdp}{q}

\newcommand{\qimdpprime}{\qimdp^\prime}

\newcommand{\pathimdp}{\omega_{\I}}
\newcommand{\pathpimdp}{\path_\PIMDPS}

\newcommand{\pathimdpfin}{\pathimdp^{\mathrm{fin}}}
\newcommand{\pathpimdpfin}{\pathpimdp^{\mathrm{fin}}}

\newcommand{\Pathimdp}{\mathit{Paths}}
\newcommand{\Pathimdpfin}{\mathit{Paths}^{\mathrm{fin}}}
\newcommand{\last}{\mathit{last}}

\newif\ifarxiv
\arxivtrue
\ifarxiv
    \usepackage{fancyhdr}
    \fancypagestyle{arxiv}{%
      \fancyhf{} 
      
      \fancyhead[C]{\vspace{-25pt}\textbf{Updated from version in the 2021 Proceedings of the IEEE Conference on Decision and Control}}
    }%
\fi

\begin{document}

\maketitle

\ifarxiv  \thispagestyle{arxiv}\pagestyle{arxiv}
\else \thispagestyle{empty}\pagestyle{empty} \fi

\begin{abstract}
Autonomous systems often have complex and possibly unknown dynamics due to, e.g., black-box components.
This leads to unpredictable behaviors and makes control design with performance guarantees a major challenge. 
This paper presents a data-driven control synthesis framework for such systems subject to linear temporal logic on finite traces (\LTLf) specifications.  The framework combines a baseline (offline) controller with a novel online controller and refinement procedure 
that improves the baseline guarantees as new data is collected.
The baseline controller is computed offline on an uncertain abstraction constructed using Gaussian process (GP) regression 
on a given dataset.
The offline controller provides a lower bound on the probability of satisfying the \LTLf specification, which may be far from optimal due to both discretization and regression errors.
The synergy arises from the online controller 
using the offline abstraction 
along with the current state and new data to choose the next best action.  
The online controller may improve the baseline guarantees since it avoids the discretization error and reduces regression error as new data is collected.  
The new data are also used to refine the abstraction and offline controller using local GP regression, which significantly reduces the computation overhead. 
Evaluations show the efficacy of the proposed offline-online framework, especially when compared against the offline controller.

\end{abstract}

\section{INTRODUCTION}
    \label{sec:intro}

The accelerating development of autonomous systems technology has led to their proliferation
in many facets of our society.
Examples include manufacturing robots on factory floors, autonomous cars on the roads, and delivery drones in urban airspaces.  Such systems often have complex and possibly unknown dynamics due to, e.g., black-box components, making accurate modeling unattainable.  This introduces a major challenge for analysis and control design for such systems.  This challenge is especially important to confront in \emph{safety-critical} domains, where assurances are required.  This work considers this challenge and aims to provide a control synthesis framework that uses data in lieu of a system model to provide performance guarantees. 

A powerful approach to control design with assurances is formal synthesis, e.g.,  \cite{tabuada2009verification,Belta:Book:2017,doyen2018verification,Lahijanian:TAC:2015,laurenti2020formal}.  In this approach, the system specification is expressed in a formal language such as \textit{linear temporal logic} (\LTL) \cite{baier2008principles}, and the system evolution is abstracted to a finite model--called \textit{abstraction}--with a simulation relation.  Then, using automated verification techniques, a controller is synthesized on the abstraction with performance guarantees.  The controller and its guarantees are then refined onto the underlying system.  These frameworks provide strong assurances but rely on accurate dynamics model to construct the abstraction, which may be unavailable for complex autonomous systems.

In recent years, data-driven control approaches have been increasingly studied, e.g., \cite{jagtap2020control, berkenkamp2017safe,Berkenkamp2016,devonport2020bayesian,maiworm2021online}.  These methods typically consist of a \emph{machine learning} component that constructs a 
a variety of regression models, e.g., polynomial functions and neural networks.
Among them, \textit{Gaussian process} (GP) regression \cite{rasmussen2006gps4ml} offers a unique advantage of providing quantified error bounds between the regressed model and the underlying system \cite{srinivas2012information,Germain2016pac,chowdhury2017kernelized, lederer2019uniform}.  This has given rise to a wide use of GPs in safe learning frameworks \cite{berkenkamp2017safe,Berkenkamp2016,devonport2020bayesian}.  In particular, safe reinforcement learning approaches are developed to take advantage of the GP modeling to learn a control policy with guarantees such as stability \cite{berkenkamp2017safe,maiworm2021online} and respecting safety constraints \cite{Berkenkamp2016, devonport2020bayesian}.
A shortcoming of classical GP regression, however, is that it leads to computational intractability as the data grows, making it difficult to use for online refinement of policies.

Sparse GP approximation has been explored extensively to handle large datasets \cite{liu2020gaussian}.
These approximations range from choosing subset of data, to making conditional assumptions on the joint prior using inducing datapoints.
While powerful, sparse GP approximations can lack analytical guarantees 
\cite{quinonero2005unifying}.
The use of localized GP regressions to model local dynamics has been studied for model learning and control~\cite{nguyen2008local,capone2020localized},
and it is shown that they can outperform sparse GP approximations if intelligent data partitioning is used~\cite{lederer2020dividing}.
Although they have been used in various control settings, formal online synthesis using GPs remains an open challenge.

In previous work \cite{jackson2021synthesis}, we introduce a data-driven formal control synthesis framework for unknown stochastic systems.  The framework takes a set of input-output data of the system and a desired property in \textit{\LTL on finite traces} (\LTLf) \cite{de2013linear} as input and synthesizes a \emph{robust} control policy with guaranteed lower-bound on the probability of achieving the \LTLf property.  The method employs GP regression and a discretization to construct an abstraction in the form of an uncertain Markov process for the unknown system with a simulation relation.  It then computes a robust policy on the abstraction with a quantified bound on the error (distance) to the optimal solution.  This error is due to both the discretization and regression errors in generating the abstraction.   Given the inherent computational complexity, the framework is completely offline and cannot take advantage of runtime data to reduce its error.

In this work, we build on \cite{jackson2021synthesis,jackson2020safety} and propose a synergistic offline-online framework to approach an optimal solution.  We combine the baseline offline controller in \cite{jackson2021synthesis} with a novel online controller and refinement procedure that improves the baseline guarantees as new data is collected.  The synergy arises from the online controller using the offline abstraction along with the current state and new data to choose the next best action.  The online controller improves the baseline guarantees by avoiding the discretization error and reducing regression error as new data is collected.  The new data are also used to refine the abstraction and offline controller via local GP regression, which significantly reduces the computation overhead, enabling online computations.  

The contributions of this work are threefold.  First, we introduce an online procedure for control synthesis that improves offline control policy's probability of satisfaction of the \LTLf specification.  Specifically, we show that the online method works in synergy with the offline framework to monotonically decrease the error to the optimal solution at runtime.  This is more general than the safety problem considered in \cite{Berkenkamp2016, devonport2020bayesian}.
Second, we reduce computational overhead to enable refinement of the abstraction online, and hence, improving the baseline controller via local GPs.  We show that a global GP is too expensive to use online, and local GPs enable comparatively quick computations.  Thirdly, we compare the performance of the offline method against the proposed synergistic framework in simulation, illustrating the efficacy of the proposed method.

\section{PROBLEM FORMULATION}
    \label{sec:problem}

We consider a stochastic system given by
\begin{equation}\label{eq:sys}
\x_{k+1} = f(\x_k, \u_k) +  \procnoise_k
\end{equation}
\noindent
where $k\in \naturals$, $\x_k \in \reals^n$, and $\u_k \in U=\{u_1, \ldots, u_{|U|} \}$ is a finite set of controls or actions.
$\x_k$ is a discrete-time controlled stochastic process governed by a fully- or partially-unknown, non-linear function $f:\reals^n\times U\to \reals^n $ driven by an additive noise term $\procnoise_k \in \reals^n$.  
We assume that $\procnoise_k$ follows a stationary and independent sub-Gaussian distribution $p_\procnoise$. 
This includes Gaussian distributions as well as all distributions with bounded support \cite{massart2007concentration}.

\newcommand{\system}{Process \eqref{eq:sys}}

To reason about Process~\eqref{eq:sys} without having full knowledge of $f$, we assume to have a set of state-action-state measurements $\dataset=\{(x_i,u_i,x^+_i)_{i=1}^m\}$ generated by Process~\eqref{eq:sys}, where $x^+_i\in\reals^n$ is a sample of one-step evolution of Process~\eqref{eq:sys} initialized at $x_i\in\reals^n$ with control $u_i\in U$.  
Our goal is to use $\dataset$ as well as the data collected at runtime to infer $f(\cdot,u)$ for each $u\in U$. The following assumption guarantees that $f(\cdot,u)$ can be learned arbitrarily well via GP regression.
\begin{assumption}
    \label{assump:smoothnes}
    For a compact set $X \subset \mathbb{R}^n$, let $\kernel:\mathbb{R}^n\times \mathbb{R}^n\to \mathbb{R}$ be a given kernel and $\mathcal{H}_\kernel(\FullSet)$ the reproducing kernel Hilbert space (RKHS) of functions over $\FullSet$ corresponding to $\kernel$ with norm $\| \cdot \|_\kernel$ \cite{srinivas2012information}.  Then, for each $u \in U$  and $i\in \{1,...,n \},$ $f^{(i)}(\cdot,u) \in \mathcal{H}_\kernel(\FullSet)$ and for a constant $B_i>0,$ it holds that  $\| f^{(i)}(\cdot,u) \|_\kernel \leq B_i$, where $f^{(i)}$ is the $i$-th component of $f$.
\end{assumption}
\noindent
For instance, assuming that $\kernel$ is the widely used squared exponential kernel (as in our experiments), then $\mathcal{H}_\kernel(\FullSet)$ is a space of functions that is dense with respect to the set of continuous functions on $X$, i.e., members of $\mathcal{H}_\kernel(\FullSet)$ can approximate any continuous function on $X$ arbitrarily well \cite{steinwart2001influence}.

We denote by $\pathX = x_0 \xrightarrow{u_0} x_1 \xrightarrow{u_1}  \ldots $ a \textit{path} or \textit{trajectory} of $\x_k$ and use $\pathX(k)=x_{k}$ to indicate the state of $\pathX$ at time $k$.  
Further, we denote by $\PathFin$ the set of all sample paths with finite length, i.e, the set of trajectories $\pathX^{k}= x_0 \xrightarrow{u_0} x_1 \xrightarrow{u_1}  \ldots \xrightarrow{u_{k-1}} x_k$ for all $k\in \naturals$. With a slight abuse of notation, given a path $\pathX$, we denote by $\pathX^k$ the prefix of $\pathX$ up to step $k$.  
A \textit{control strategy} $\policy_\x: \PathFin \to U$ is a function that chooses the next control $u \in U$ given a finite path $\pathX^{k}\in \PathFin$. 

For $u\in U$, 
a Borel measurable set $\FullSet\subseteq \reals^n$, and $x\in \reals^n,$ 
call 
$$T(\FullSet \mid x,u)=\int \mathbf{1}_\FullSet(f(x,u)+\bar{v})\distribution_{\procnoise}(\bar v)d \bar v,$$ 
the \emph{stochastic transition function}  induced by Process \eqref{eq:sys}, where 
\begin{equation*}
    \mathbf{1}_\FullSet(x) =
    \begin{cases}
        1 & \text{if } x\in \FullSet\\
        0 & \text{otherwise}
    \end{cases}
\end{equation*}
is the indicator function. 
Kernel $T(\FullSet \mid x,u)$ describes the probability of $\x$ ending in set $X$ in one-step evolution given the current state $x$ and control $u$.
Given a control strategy $\policy_\x$ and a time horizon $[0,N],$ it is possible \cite{bertsekas2004stochastic} to define a probability measure $P$ over the paths of $\x_k$ uniquely generated by $T$  and a (fixed) initial condition $x_0\in \mathbb{R}^n$ 
such that 
$$P[\pathX^N(0)\in \FullSet] = \mathbf{1}_{\FullSet}(x_0),$$ and for $k\in \{1,...,N\}$,
\begin{equation*}
    P[\pathX^N(k) \in \FullSet \mid \pathX^N(k-1)=x, \policy_\x] = T(\FullSet \mid x, {\policy_\x(\pathX^{k-1})}).
\end{equation*}
Furthermore, for $N=\infty$, $P$ is still uniquely defined by $T$ by the \emph{Ionescu-Tulcea extension theorem} \cite{abate2014effect}. 

\subsection{Specification Language}
\label{sec:ltlf}
We are interested in the properties of Process~\eqref{eq:sys} in a compact set $\FullSet \subset \reals^n$ with respect to a finite set of closed regions of interest $\ROISet = \{\roi_1,\ldots,\roi_{|\ROISet|}\}$, where $\roi_i \subseteq \FullSet$.  To this end, we associate to each region $\roi_i$ the atomic proposition $\Prop_i$ such that $\Prop_i = \top$ (i.e., $\Prop_i$ is \textit{true}) if $x \in \roi_i$; otherwise $\Prop_i = \bot$ (i.e., $\Prop_i$ is \textit{false}).  
Let $\APs = \{\Prop_1, \ldots,\Prop_{|\ROISet|}\}$ denote the set of all atomic propositions and $L: \FullSet \rightarrow 2^{\APs}$ be the labeling function that assigns to state $x$ the set of atomic propositions that are true at $x$.
Then, we define the \textit{trace} of path 
$\pathX^{k}= x_0 \xrightarrow{u_0} x_1 \xrightarrow{u_1}  \ldots \xrightarrow{u_{k-1}} x_k$
to be
\begin{equation*}
	\trace = \trace_0 \trace_1 \ldots \trace_k,
\end{equation*}
where $\trace_i = L(x_i) \in 2^{\APs}$ for all $i \leq k$. 
With an abuse of notation, we use $L(\pathX^k)$ to denote the trace of $\pathX^k$.

We use a temporal logic to express desired properties of Process~\eqref{eq:sys}.  Classical temporal logics such as \LTL  specifications \cite{baier2008principles}, however, are interpreted over infinite behaviors (traces), and given the high levels of uncertainty of Process~\eqref{eq:sys} (unknown dynamics as well as noise), its infinite behaviors have trivial probabilities.  Therefore, we instead employ recently developed \textit{\LTL over finite traces} (\LTLf) \cite{de2013linear}, which has the same syntax as \LTL  but its semantics is defined over finite traces. 

\begin{definition}[\LTLf Syntax]
    An \LTLf formula $\Spec$ over a set of atomic propositions $\APs$ is recursively defined as
	\begin{equation*}
		\Spec := \top \mid \Prop   \mid   \neg \Spec   \mid   \Spec \wedge \Spec   \mid  \Next \Spec  \mid  \Spec\, \mathcal{U} \Spec
	\end{equation*}
	where $\Prop \in \APs$, $\neg$ (negation) and $\wedge$ (conjunction) are Boolean operators, and $\Next$ (next) and $\mathcal{U}$ (until) are temporal operators.
\end{definition}

\begin{definition}[\LTLf Semantics]
	The semantics of an \LTLf formula $\Spec$ are defined over finite traces. 
	Let trace $\rho \in (2^{\APs})^*$, $|\trace|$ denote its length, and $\trace_i$ be the $i$-th symbol of $\trace$.  
	Then, the satisfaction of $\Spec$ by the $i$-th step of $\trace$, denoted by $\trace, i \models \Spec$, is recursively defined as
	\begin{itemize}
		\item $\trace , i \models \top,$
		\item $\trace , i \models \Prop \quad  \Leftrightarrow  \quad \Prop \in \trace_i,$
		\item $\trace , i \models \neg \Spec \quad \Leftrightarrow \quad \trace, i \not \models \Spec,$
		\item $\trace , i \models \Spec_1 \wedge \Spec_2 \quad \Leftrightarrow \quad \trace, i \models \Spec_1$ and $\trace, i \models \Spec_2,$
		\item $\trace , i \models \Next \Spec \quad \Leftrightarrow \quad |\trace| > i+1$ and $\trace, i+1 \models \Spec,$
		\item $\trace , i \models \Spec_1 \mathcal{U} \Spec_2 \quad \Leftrightarrow \quad \exists j$ s.t. $ i \leq j < |\trace|$ and $\trace, j \models \Spec_2$ and $\forall k$, $i \leq k < j$, \; $\trace, k \models \Spec_1.$
	\end{itemize}
	Finite trace $\trace$ satisfies $\Spec$, denoted by $\trace \models \Spec$, if $\trace, 0 \models \Spec$.
\end{definition}
\noindent
Similar to \LTL, the temporal operators $\Eventually$ (eventually) and $\Globally$ (globally) are defined as: 
$\Eventually \, \Spec = \top \, \mathcal{U} \, \Spec$ and $\Globally \, \Spec = \neg \Eventually \, \neg \Spec.$

An \LTLf formula $\Spec$ defines a language $\mathcal{L}(\Spec) = \{\trace \in (2^{\APs})^* \mid \trace \models \Spec\}$, which is in fact a regular language.  
Similar to \cite{wells2020ltlf}, we say that a path $\pathX$ of Process~\eqref{eq:sys} satisfies \LTLf formula $\Spec$ if there exists a prefix of $\pathX$ that lies entirely in $\FullSet$ and its trace is in  $\mathcal{L}(\Spec)$, i.e., 
\begin{multline}
    \label{eq:satisfying-path}
    \pathX \models \Spec  \quad \Leftrightarrow \quad \exists k \in \naturals  \;\; s.t. \;\; L(\pathX^k) \in \mathcal{L}(\Spec) \text{ and } \\ \pathX^k(k') \in \FullSet \;\; \forall k' \leq k,
\end{multline}

\subsection{Problem Statement}

Given an \LTLf specification $\Spec$ and dataset $\dataset$, our goal is to synthesize a control strategy $\optpolicy_\x$ under which Process \eqref{eq:sys} attempts to satisfy $\Spec$ with the maximum probability.

\begin{problem}\label{prob:main}{[Offline-Online Synthesis]}
Given an initial dataset $\dataset$ and the capability of collecting data of Process \eqref{eq:sys} at runtime, a compact set $X$, 
and
an \LTLf property $\Spec$ defined over the regions of interest in $X$, find a control strategy  $\optpolicy_{\x}$ 
that maximizes the probability of satisfying $\Spec$, i.e.,
$$ \optpolicy_{\x} = \arg \max_{\policy_\x} P[\pathX \models \Spec \mid \policy_\x, \pathX(0)=x_0].$$
\end{problem}

The fact that  Process \eqref{eq:sys} is unknown makes solving Problem \ref{prob:main} particularly challenging.  In this work, we develop a data-driven method that achieves the optimality objective of the problem at the limit.
In particular, we develop an offline-online synergistic framework, where the offline module uses GP regression on $\dataset$ to build an initial abstraction of Process \eqref{eq:sys} in terms of a finite uncertain Markov model and synthesizes a robust control strategy on the resulting abstraction.  
This abstraction and control strategy suffer from two types of errors: discretization and regression errors.  
Via the online module, we iteratively reduce these errors as explained below.  

The offline strategy serves as a baseline controller for the online module and provides an initial performance guarantee in terms of the lower bound on the probability of satisfying $\Spec$.
Online, while executing the system, we use the observed data to refine the abstraction, potentially improving the control strategy and its probability of satisfaction of $\Spec$ in two ways: 1) we reduce the space discretization error in the abstraction by reducing the uncertainty in the transition probabilities of the abstraction, which can be done by leveraging our continuous state knowledge at every time step, and 2) using the data collected at each time step, we improve our learned model of Process~\eqref{eq:sys}, hence, reducing the regression error. 
We show that updates to the abstraction and baseline controller at runtime is computationally feasible by learning a series of local GPs instead of a single global GP.  Furthermore, we show that the framework formally accounts for the uncertainty in the learning process, and the online guarantees see monotonic improvement over those generated offline.


\section{PRELIMINARIES}
    \label{sec:prelim}
\subsection{Gaussian Process Regression}
{Gaussian Process} (GP) regression is a non-parametric Bayesian machine learning method \cite{rasmussen2006gps4ml} that aims to infer an unknown function $f:\mathbb{R}^n\to \mathbb{R}$ from noisy data.
A standard assumption of GP regression is that $f$ is a sample from a GP with zero mean and covariance $\kernel: \mathbb{R}^n \times \mathbb{R}^n \to \mathbb{R}$. 
Let $\dataset=\{(\mathrm{x}_i,\mathrm{y}_i),i\in \{1,\dots,m \}\}$ be a dataset, where 
$\mathrm{y}_i$ is a sample of an observation of $f(\mathrm{x}_i)$ with independent zero-mean noise ${v}$, which is assumed to be normally distributed with variance $\sigma^2$ and  $\mathrm{X}$ and $\mathrm{Y}$ be ordered vectors with all points in $\dataset$ such that $\mathrm{X}_i = \mathrm{x}_i$ and $\mathrm{Y}_i = \mathrm{y}_i$.  Further, call $K(\mathrm{X},\mathrm{X})$ the matrix with $K_{i,j}(\mathrm{X}_i,\mathrm{X}_j)=\kernel(\mathrm{x}_i,\mathrm{x}_j)$, $K(\mathrm{x},\mathrm{X})$ the vector such that $K_{i}(\mathrm{x},\mathrm{X})=\kernel(\mathrm{x},\mathrm{X}_i)$, and $K(\mathrm{X},\mathrm{x})$ defined accordingly. Then, the predictive distribution of $f$ at a test point $\mathrm{x}$ is given by the conditional distribution of $f$ 
given $\dataset$, which is Gaussian with mean $\mu_\dataset$ and variance $\sigma_\dataset^2$ given by
\begin{align}
    \begin{split}\label{eq:post-mean}
      & \mu_{\dataset}(\mathrm{x}) = K(\mathrm{x},\mathrm{X}) \big( K(\mathrm{X},\mathrm{X})+ \sigma^2I_{m} \big)^{-1} Y   \\
    \end{split}\\
    \begin{split}\label{eq:post-kernel}
      & \sigma_{\dataset}^2(\mathrm{x}) = \kernel(\mathrm{x},\mathrm{x})- K(\mathrm{x},\mathrm{X})\big( K(\mathrm{X},\mathrm{X})+ \sigma^2 I_{m} \big)^{-1}K(\mathrm{X},\mathrm{x}),
     \end{split}
\end{align}
where $I_{m}$ is the identity matrix of size $m \times m$. 
For stationary kernels, such as the squared-exponential kernel, $\sigma_\dataset(\cdot)$ sees monotonic decay as $m$ increases.  
As $m\to\infty$, the posterior covariance $\sigma_\dataset(\cdot)\to 0$ and $\mu_{\dataset}\to g$~\cite{rasmussen2006gps4ml}.

In this work, contrary to the standard GP regression setting described above, we do not assume that $f$ is sampled from a given GP or that the observation noise is Gaussian. To quantify the estimation error in this more agnostic setting, under Assumption \ref{assump:smoothnes}, we can rely on Theorem 2 from \cite{chowdhury2017kernelized}, which bounds the regression error with high-probability in the form:
\begin{equation}
    \label{eq:GPerror}
    P\big[\forall \mathrm{x} \in \FullSet, \; |\mean_\dataset(\mathrm{x}) - f(\mathrm{x})| < \beta(\delta)\sigma_{\dataset}(\mathrm{x}) \big]\geq 1-\delta,
\end{equation}
where $\delta\in (0,1)$ specifies the upper bound, and $\beta$ is a compensator that depends on the choice of $\kernel$, the dataset $\dataset$, and $\delta$. We refer to \cite{chowdhury2017kernelized} for further information on $\beta$ and \eqref{eq:GPerror}.  

\subsection{Interval Markov Decision Processes (IMDPs)}
 We use \textit{interval Markov decision process} (IMDPs) as the abstraction model for Process \eqref{eq:sys}. IMDPs generalize Markov decision processes (MDPs) by allowing an interval of values for transition probabilities \cite{givan2000bounded}.  

\begin{definition}[{\imdp}] \label{def:imdp}
    An interval Markov decision process ({\imdp}) is a tuple $\I = (\Qimdp,\Aimdp,\Plow,\Pup,\APs,L)$, where
    \begin{itemize}
    	\setlength\itemsep{1mm}
    	\item $\Qimdp$ is a finite set of states,
    	\item $\Aimdp$ is a finite set of actions, and $\Aimdp(\qimdp)$ denotes the set of available actions at state $\qimdp \in \Qimdp$,
        \item $\Plow: \Qimdp \times \Aimdp \times \Qimdp \rightarrow [0,1]$ is a function, where $\Plow(\qimdp,a,\qimdpprime)$ defines the lower bound of the transition probability from state $\qimdp \in \Qimdp$ to state $\qimdpprime \in \Qimdp$ under action $a \in \Aimdp(\qimdp)$,
        \item $\Pup: \Qimdp \times \Aimdp \times \Qimdp \rightarrow [0,1]$ is a function, where $\Pup(\qimdp,a,\qimdpprime)$ defines the upper bound of the transition probability from state $\qimdp \in \Qimdp$ to state $\qimdpprime \in \Qimdp$ under action $a \in \Aimdp(\qimdp)$,
        \item $\APs$ is a finite set of atomic propositions,
        \item $L: \Qimdp \rightarrow 2^{\APs}$ is a labeling function that assigns to each state $\qimdp \in \Qimdp$ a subset of $\APs$.
    \end{itemize}
\end{definition}
For all $\qimdp,\qimdpprime \in \Qimdp$ and $a \in \Aimdp(\qimdp)$, it holds that $\Plow(\qimdp,a,\qimdpprime) \leq \Pup(\qimdp,a,\qimdpprime)$ and $
    \sum_{\qimdpprime \in \Qimdp} \Plow(\qimdp,a,\qimdpprime) \leq 1 \leq \sum_{\qimdpprime \in \Qimdp} \Pup(\qimdp,a,\qimdpprime)$.
 
A path of an \IMDPText is a sequence of states $\pathimdp = \qmdp_0 \xrightarrow{\amdp_0} \qmdp_1 \xrightarrow{\amdp_1} \qmdp_2 \xrightarrow{\amdp_2}  \ldots$ such that $\amdp_k \in \Amdp(\qmdp_k)$ and $\Pup(\qmdp_k, \amdp_k, \allowbreak{\qimdp_{k+1}}) > 0$ for all $k \in \naturals$. We denote 
the last state of a finite path $\pathimdpfin$ by $\last(\pathimdpfin)$ and the set of all finite and infinite paths by $\Pathimdpfin$ and $\Pathimdp$, respectively. 
\begin{definition}[Strategy]
\label{def:strategy}
    A strategy $\policy$ of an \imdp model $\I$ is
    a function $\policy: \Pathimdpfin \rightarrow \Aimdp$ that maps a finite path $\pathimdpfin$ of $\I$ onto an action in $\Aimdp(\last(\Pathimdpfin))$. If a strategy depends only on $\last(\pathimdpfin)$, it is called a memoryless (stationary) strategy.  
    The set of all strategies is denoted by $\Setofpolicies$.
\end{definition}

Once an action is chosen by strategy $\policy$, we evolve from the current state to the next state according to a probability distribution that respects the transition probability bounds of the \IMDPText.  There exist possibly infinitely many such distributions, and an \textit{adversary} chooses this distribution.
\begin{definition}[Adversary]
\label{def:adversary}
    For an \imdp $\I$, an adversary is a function $\adversary: \Pathimdpfin \times \Aimdp \rightarrow \probDist(\Qimdp)$ that, for each finite path $\pathimdpfin \in \Pathimdpfin$, state $\qimdp=\last(\pathimdpfin)$, and action $a \in \Aimdp(\last(\pathimdpfin))$, assigns a feasible distribution $\feasibleDist{\qimdp}{a}$ which satisfies
    $\Plow(\qimdp,a,\qimdpprime) \leq \feasibleDist{\qimdp}{a}(\qimdpprime) \leq \Pup(\qimdp,a,\qimdpprime).$
    The set of all adversaries is denoted by $\Adversary$.
\end{definition}



\noindent
Given a strategy $\policy$ and an adversary $\adversary$, a probability measure over the paths of \IMDPText $\I$ can be defined using the induced Markov chain \cite{Lahijanian:TAC:2015}.

\subsection{Deterministic Finite Automaton (DFA)}
Given \LTLf formula $\Spec$, a DFA can be constructed that precisely accepts the language of $\Spec$ \cite{de2013linear}.
\begin{definition}[DFA]
    A \textit{deterministic finite automaton} (DFA) constructed from an \LTLf formula $\Spec$ defined over atomic propositions $\APs$ is a tuple $\DFAA=(\DFAStates, 2^{\APs}, \DFATransition, \DFAState_0,\DFAFinalState)$, where 
    $\DFAStates$ is a finite set of states,
    $2^{\APs}$ is a finite set of input symbols,
    $\DFATransition:\DFAStates\times 2^{\APs} \to\DFAStates$ is the transition function, 
    $\DFAState_0 \in \DFAStates$ is the initial state, and
    $\DFAFinalState \subseteq \DFAStates$ is the set of accepting (final) states. 
\end{definition} 
A finite \emph{run} on $\DFAA$ is a sequence of states $\DFAState_0\DFAState_1\dots\DFAState_{n+1}$ induced by trace $\trace = \trace_0\trace_1\dots\trace_n$, where $\trace_i \in 2^{\APs}$ and $\DFAState_{i+1} = \DFATransition(\DFAState_i,\trace_i)$. A finite run on $\trace$ is accepting if $\DFAState_{n}\in\DFAFinalState$.  If the run is accepting, then trace $\trace$ is accepted by $\DFAA$.  The set of all traces that are accepted by $\DFAA$ is call the language of $\DFAA$, denoted by $\mathcal{L}(\DFAA)$.  This language is equal to the language of $\Spec$, i.e., $\mathcal{L}(\Spec) = \mathcal{L}(\DFAA)$. 
    
\section{OFFLINE-ONLINE SYNTHESIS}
    \label{sec:synthesis}

In this section, we describe our synergistic offline-online control synthesis framework, which consists of two modules.  The offline module constructs an IMDP abstraction for Process~\eqref{eq:sys} and generates the baseline control strategy with a lower-bound guarantee on the probably of satisfying $\Spec$.  The online module uses the offline abstraction and lower-bound probabilities to refine the control strategy and abstraction, which results in monotonic improvements to the guarantees.  
We first briefly describe the offline module, which is based on previous work \cite{jackson2021synthesis}, and then detail the online module, which is the main contribution of this work.  

\subsection{Offline Module}
\label{sec:offline}

The offline module first performs GP regression to estimate the unknown function $f(\cdot,u)$ for each $u \in U$ from the given initial dataset $\dataset$.  This is achieved by using \eqref{eq:post-mean} and \eqref{eq:post-kernel} on each component of $f$.  Note that the corresponding regression error $|\gest^{(i)}(x,u) - f^{(i)}(x,u)|$, where $\gest^{(i)}$ is the estimate (GP mean) of the $i$-th component of $f$ obtained from \eqref{eq:post-mean}, 
is given by \eqref{eq:GPerror}.  

Given $\gest$, the method constructs the IMDP abstraction $\I = (\Qimdp, \Aimdp ,\Plow, \Pup, \APs, L)$ by partitioning of the compact set $X$.  Each resulting discrete region is associated with an IMDP state  $\qimdp \in \Qimdp$. 
One state of the IMDP is also used to represent the rest of the continuous space $\reals^n \setminus X$.
The labeling function $L$ of the IMDP states is defined according to the labels of continuous states within each region.  We remark that, to ensure correct labeling, the space discretization performed above must respect the regions of interest in $\ROISet$.  This guarantees that the label of the discrete regions correctly hold for every continuous state in that region, i.e., for discrete region $q \subset \reals^n$, $L(q) = L(x)$ for every $x \in q$.  

Next, the set of IMDP actions is defined to be the set of controls of Process~\eqref{eq:sys}, i.e., $\Aimdp = U$.  The transition probability bounds are then given by the following proposition, which uses the discrete regions, the regressed GP, and its corresponding error.

\begin{proposition}[\hspace{-1.2mm} \cite{jackson2021synthesis}, Theorem 1]\label{prop:transition}
    For a region $q \subset \reals^n$ and control $\aimdp \in \Aimdp$, let $\|h\|^{\qimdp,\aimdp}_{\infty}\equiv \sup_{x\in\qimdp} |h(x,u)|$ .
    Given a control $\aimdp \in\Aimdp$, regions $\qimdp,\qimdp' \subset \reals^n$, dataset $\dataset$, regression $\gest$, and 
    positive real vectors $\distBound\in\reals^n$ and $\eta\in\reals^n$, then
    \begin{align}
        \Pup&(\qimdp,\aimdp,\qimdp') = \max_{x\in \qimdp} T(\qimdp' \mid x, \aimdp) \nonumber \\ 
            &\leq \big (\interindicator{\overline {\qimdp}' (\distBound+\eta)}{ Im(\qimdp,\aimdp)} 
            \prod_{i=1}^n P[|w^{(i)}|\leq \eta_i] + \prod_{i=1}^n P[|w^{(i)}|>\eta_i]\big)~\cdot  \nonumber \\ 
            & \prod_{i=1}^n P[\|\gest^{(i)}-f^{(i)}\|^{\qimdp,\aimdp}_\infty \leq \distBound_i] + \prod_{i=1}^nP[\|\gest^{(i)} - f^{(i)}\|^{\qimdp,\aimdp}_\infty >\distBound_i],
            \label{eq:upper-bound}
    \end{align}
    \begin{align}
        \Plow&(\qimdp,\aimdp,\qimdp') = \min_{x\in \qimdp} T(\qimdp'\mid x,\aimdp)  \nonumber \\
            & \geq \big(1-\interindicator{\FullSet \setminus \underline {\qimdp}'(\distBound+\eta)}{Im(\qimdp,\aimdp)}\big)  \prod_{i=1}^n P[\|\gest^{(i)}-f^{(i)}\|^{\qimdp,\aimdp}_\infty \leq \distBound_i]~\cdot \hspace{25mm} \nonumber \\
            & \hspace{48mm} \prod_{i=1}^n P[|w^{(i)}|\leq \eta_i].
        \label{eq:lower-bound}
    \end{align}
    where $\interindicator{A}{B}$ returns one if $A\cap B\neq\emptyset$ otherwise zero, $\overline{\qimdp}'(c)$ 
    and $\underline {\qimdp}'(c)$ are the regions obtained by expanding and shrinking (each dimension of) $\qimdp'$ by the scalars in vector $c$, respectively, and $Im(\qimdp,\aimdp)$ is the image of $\qimdp$ under $\gest$ defined as $Im(\qimdp,\aimdp) = \{\gest(x,\aimdp) \mid x\in\qimdp\}$. 
\end{proposition}
Further details of Proposition~\ref{prop:transition} can be found in \cite{jackson2021synthesis}. 
Note that the discretization (volume of $\qimdp$ and $\qimdp'$) plays a significant role in the transition probability bounds as the image of $\qimdp$ under the learned dynamics can be very large (especially when it is over-approximated).
In addition, the regression errors dictate both the $P[\|\gest^{(i)}- f^{(i)}\|^{\qimdp,\aimdp}_\infty \leq \distBound_i]$ terms along with affecting the magnitude of the expanded and reduced sets.  Hence, the larger these errors, the looser the bounds $\Plow$ and $\Pup$, leading to larger uncertainty in the abstraction. 

Given abstraction $\I$ and specification $\Spec$, a control strategy can be generated that is robust against these errors and maximizes the lower bound probability of satisfying $\Spec$.  This is achieved via a product construction between abstraction $\I$ and \DFAText $\DFAA$, which represents $\Spec$.
The result is the product \IMDPText (\PIMDPText) which is defined below.


\begin{definition}[Product \IMDPText]\label{def:pimdp}
    Given an \IMDPText $\IMDP = (\Qimdp, \Aimdp, \Plow,\Pup, \APs, L)$ and \DFAText $\DFAA = (\DFAStates, 2^{\APs}, \DFATransition, \DFAState_0,\DFAFinalState)$, the \textit{product \IMDPText} (\PIMDPText) $\PIMDP 
    =(\Qpimdp, \Aimdp, \Plow^{\PIMDP}, \Pup^{\PIMDP}, \Qpimdp_0, \Qpimdp_F)$, where 
    $\Qpimdp = \Qimdp\times\DFAStates$, $\Qpimdp_F = \Qimdp\times\DFAStates_F,$
    $$\Qpimdp_0 = \{(\qimdp,\DFAState_{init}) \mid  \qimdp \in \Qimdp , \; \DFAState_{init}=\DFATransition(\DFAState_0, L(\qimdp))\},$$
    and
    \begin{align*}
        \Plow^{\PIMDP}((\qimdp, \DFAState), \aimdp, (\qimdp', \DFAState')) &= \begin{cases}
            \Plow(\qimdp,\aimdp,\qimdp') & \text { if } \DFAState'=\DFATransition(\DFAState,L(\qimdp))\\
            0 & \text{ otherwise}
        \end{cases} \\
        \Pup^{\PIMDP}((\qimdp, \DFAState), \aimdp, (\qimdp', \DFAState')) &= \begin{cases}            \Pup(\qimdp,\aimdp,\qimdp') & \text { if } \DFAState'=\DFATransition(\DFAState,L(\qimdp))\\
            0 & \text{ otherwise}.
        \end{cases} 
    \end{align*}
\end{definition}
The $\PIMDPText$ accepting states $\Qpimdp_F$ encapsulate the satisfaction of the specification $\spec$, i.e., a path that reaches $\Qpimdp_F$ satisfies $\spec$.
Hence, we formulate the following optimization problem to compute a robust strategy $\policyoff:\Qpimdp\to\Apimdp$ that maximizes the lower bound of the probability of reaching a state in $\Qpimdp_F$:
\begin{equation}
    \label{eq:strategy-offline}
    \policyoff(\qpimdp) = \arg\max_{\policy \in \Setofpolicies} \;\, \min_{\adversary \in \Adversary} P[\pathpimdp \models \Spec \mid \policy, \pathpimdp(0) = \qpimdp].
\end{equation}
We solve this optimization problem using an interval-value iteration as detailed in \cite{Lahijanian:TAC:2015}, which results in a memoryless strategy on $\PIMDP$.  The interval-value iteration procedure also returns the lower- and upper-bounds of the probability of satisfaction of $\Spec$ under $\policyoff$, i.e., 
\begin{align}
    \label{eq:plow}
    \plow_{\policyoff}(\qpimdp) &= \min_{\adversary \in \Adversary} P[\pathpimdp \models \Spec \mid \policyoff, \pathpimdp(0) = \qpimdp], \\
    \pupp_{\policyoff}(\qpimdp) &= \max_{\adversary \in \Adversary} P[\pathpimdp \models \Spec \mid \policyoff, \pathpimdp(0) = \qpimdp],
    \label{eq:pup}
\end{align}
respectively.  
These satisfaction intervals let us identify regions where success is more likely (a lower-bound near one), regions where violation is more likely (an upper-bound near zero), and regions with 
high uncertainty (large intervals).  By deploying the system with $\policyoff$ at $\qimdp$, the system is guaranteed to satisfy $\spec$ with a probability at least  $\plow_{\policyoff}((\qimdp,\DFAState_{init}))$.  The goal of the online module is to synthesize controllers at runtime to improve this guarantee towards the optimal probability, which is evidently most helpful for the regions with high uncertainty. 



\subsection{Online Module}\label{sec:online}

In constructing the \PIMDPText $\PIMDP$ and strategy $\policyoff$ offline, we are limited by the uncertainty introduced by the finite amount of data in $\dataset$ and by the discretization of $\FullSet$. Our goal is to use the information collected at runtime to improve $\policyoff$ by reducing both sources of uncertainty.
In Alg. \ref{alg:onlineloop}, we present a high-level description of our synergistic control framework that obtains both of these goals. 

\begin{algorithm}[b!]
\caption{Online control loop}\label{alg:onlineloop}
\begin{algorithmic}[1]
\Procedure{ControlLoop}{$\PIMDP, \GPFullSet, x_0, \dataset$}
  \State $k\gets 0$
  \State $\qpimdp_0\gets$\textsc{InitialPIMDPState}$(\PIMDP, x_0)$
 \While{$\qpimdp_k\notin\Qpimdp_F$} 
  \State $\GPLocalSet\gets$\textsc{CreateLocGP}($\GPFullSet, x_k, \dataset$)  \label{alg:line:localgp}
  \State $\PIMDP\gets$\textsc{UpdatePIMDP}$(\PIMDP, \GPLocalSet)$  \label{alg:line:pimdpupdate}
  \State $\apimdp^*\gets$
  \textsc{GetOptimalAction}$(\PIMDP, x_k)$ \label{alg:line:optimalaction}
  \State $x_{k+1}\gets f(x_k, \apimdp^*) + w_k$
  \State $\qpimdp_{k+1}\gets $\textsc{UpdatePIMPDState}$(\PIMDP, x_{k+1})$ 
  \State $\dataset\gets \dataset\cup \{(x_k,\apimdp^*,x_{k+1}) \}$
  \State $k\gets k+1$
 \EndWhile
 \EndProcedure
 \end{algorithmic}
\end{algorithm}

Online, at each time step, we observe the current state of the system $x_k$ and build a set of (local) GPs
to make predictions for the transitions of Process \eqref{eq:sys} to the states of the abstraction in one-time step starting from $x_k$ (Line 5).  Note that this set of GPs ($\GPLocalSet$) are employed exclusively to make local predictions starting from $x_k$. Hence, to train them, we can simply use only data local to $x_k,$ thus making the training process efficient enough to be employed at runtime.
The resulting GPs are then employed to refine the PIMDP $\PIMDP$ and get the optimal action $\apimdp^*$ using the values computed offline as explained below. Until we reach an accepting state of $\PIMDP$, we continue the process and keep enlarging $\dataset$ with the runtime observations.
Below, we give further details of how we update $\PIMDP$ and we select the optimal action as well as how we train local GPs $\GPLocalSet$.



\subsubsection{Optimal Actions}
To choose the optimal action, we first augment the \PIMDPText with the current state and then choose the optimal next action (Lines~\ref{alg:line:localgp} and \ref{alg:line:optimalaction} in Alg. \ref{alg:onlineloop}).
Given the current state $x_k$, we begin 
by defining a new $\PIMDP$ state
$\qpimdp' = (x_k,z)$, where $z$ is the current state of the \DFAText. By reasoning over a singleton ($x_k$ instead of region $q$ that contains $x_k$), we eliminate the discretization error over the next step. In particular, for $z'=\delta(z,L(x_k))$ we can compute 
\begin{align}
    \label{Eqn:improvedBound}
    \Plow^{\PIMDP}((x_k, \DFAState), \aimdp, (\qimdp', \DFAState'))=P[\x_{k+1}\in \qimdp' \mid \omega_{\x}(k-1)=x_k,\aimdp],
\end{align} and similarly for the upper bound.
The bound in \eqref{Eqn:improvedBound} can be computed directly using Proposition~\ref{prop:transition} and is guaranteed to improve compared to $\Plow^{\PIMDP}((\qimdp, \DFAState), \aimdp, (\qimdp', \DFAState'))$ as we can compute the image $Im(\{x_k\})$ under the learned dynamics exactly and mitigate this discretization error.

We now focus on how to choose the best action in order to reach a state in $\Qpimdp_F$. In particular, similar to the goal of the optimization problem in \eqref{eq:strategy-offline}, we want to find the action $\apimdp^*$ 
that maximizes (with an abuse of notation)
\begin{equation}\label{eq:lbsat}
    \plow(\qpimdp',\aimdp) = \min_{\adversary\in\Adversary} \sum_{\qpimdp\in \PossStates} \plow_{\policyoff}(\qpimdp)
\adversary(\pathpimdpfin, \apimdp)(\qpimdp),
\end{equation}
where $\PossStates=\{\qpimdp\in\Qpimdp\mid\Pup^{\PIMDP}(\qpimdp',\apimdp,\qpimdp) > 0\}$.
Note that $\plow(\qpimdp',\aimdp)$ is the \emph{worst-possible} expected lower-bound probability under action $\aimdp$.
We pick the best control as 
\begin{equation}\label{eq:opt-action}
    \aimdp^* = \arg \max_{\aimdp \in U} \; \plow(\qpimdp',\aimdp).
\end{equation}


In practice, it is common to have multiple actions that maximize
$\plow(\qpimdp',\apimdp)$ e.g. when it is $0$ for all actions.
We introduce secondary and tertiary metrics grounded encouraging progression, though maximizing $\plow(\qpimdp',\apimdp)$ is always preferred. 

\paragraph{Satisfaction Upper Bound}
As a secondary criteria, we consider the upper-bound of the probability of satisfaction from $\qpimdp'$, which is defined as 
\begin{equation}\label{eq:ubsat}
\pupp(\qpimdp',\apimdp) = \min_{\adversary\in\Adversary} \sum_{\qpimdp\in \PossStates} \pupp_{\policyoff}(\qpimdp)\adversary(\pathimdpfin, \apimdp)(\qpimdp).
\end{equation}
In particular, we select the action $\apimdp^*=\argmax_{\aimdp \in U^*} \pupp(\qpimdp',\apimdp),$ where $U^*$ is the set of actions returned by \eqref{eq:opt-action}. Hence, $\apimdp^*$ is the one with the greatest lower- and upper-bound probability of satisfying  $\spec.$
Note that we are again selecting actions against the worst adversary, which is required in order to be robust against uncertainty.

Even with these primary metrics defined, in practice they are not always sufficient to realize good online behavior, and there may be multiple $u^*$ optimizing both criteria.
As we are concerned with choosing $\apimdp^*$ that best improves the performance for the next step, we introduce two tertiary metrics that encourage one-step progression.

\paragraph{Sink-State Metric}
This metric prefers actions that do not violate $\Spec$ by using the inherent \DFAText sink state, from which reaching the accepting state is impossible.
For each $\qpimdp\in\PossStates$
, we check if the \DFAText transitions to the sink state and prefer
actions with a low chance of doing so. 

\paragraph{Progression Metric}
We consider a progression metric that defines progress towards reaching $\Qpimdp_F$. 
Given $\qpimdp'$, we can measure its distance from an accepting state in $\Qpimdp_F$ by using the minimum distance to $\Qpimdp_F$ from each $\qpimdp\in\PossStates$, which is easily computed offline.
Once online, we choose the action that gives the shortest expected distance to $\Qpimdp_F$.

\subsubsection{Local GP and PIMDP Updates}
At runtime, we collect data that were not available offline. 
By using this new data, we can refine the regression error and in turn obtain a less uncertain IMDP abstraction.
Rather than updating the full GP corresponding to action $\apimdp$, $\GPFull^{\apimdp}$, we construct a local GP $\GPLocal^{\apimdp}$ using the $\nns$-nearest datapoints in $\dataset$ to  $x_k$.  
Training then proceeds as for standard GP regression using the same hyperparameters as $\GPFull^{\apimdp}$.
The hyperparameter $\nns$ (number of data employed to train our local GP) induces a trade-off between regression accuracy and computation overhead. For more insight on picking $\nns$ we refer to~\cite{Nguyen-Tuong2010,das2018fast}.


Our motivation for using local GP regression stems from computational tractability in the online setting. 
Performing regression with local data is $O(\nns^3)$ in the worst case.
Thus, by choosing $\nns$ much smaller than the total number of available data, we can have a polynomial speed up that allows for runtime computations.
Note that once we have $\GPFull_{\apimdp^*}^{x_k}$, we can update the abstraction and product \PIMDPText by recalculating $\Plow(\qimdp,\aimdp^*,\qimdp')$ and $\Pup(\qimdp,\aimdp^*,\qimdp')$ for all $\qimdp$ in a neighbourhood of $x_k$. 
The transitions are then updated in $\I$ and consequently in $\PIMDP$ only if the resulting interval is tighter.
Once enough transitions are updated in the abstraction, the offline strategy $\policyoff$ can be updated using the refined \PIMDPText, denoted $\policy_k$.  
This is done again using interval-value iteration, which is polynomial time (cubic) in the size of $\PIMDP$.



\subsection{Correctness}
\noindent
The following theorem guarantees that Alg. \ref{alg:onlineloop} monotonically reduces the uncertainty in the system predictions.
\begin{theorem}
\label{th:MonotonicImprovement}
    Let $x_k$ be the state of Process \eqref{eq:sys} at the $k$-th step of Alg. \ref{alg:onlineloop}. Similarly, let $\policy_k$ be the refinement of $\policyoff$ with $k$ additional datapoints computed according to \eqref{eq:strategy-offline}. Then, for any $\qpimdp \in \Qpimdp$ it holds that 
    \begin{equation}
        [\plow_{\policy_{k+1}}(\qpimdp),\pupp_{\policy_{k+1}}(\qpimdp)]\subseteq [\plow_{\policy_k}(\qpimdp),\pupp_{\policy_k}(\qpimdp)].    
    \end{equation}
\end{theorem}
The proof of Theorem \ref{th:MonotonicImprovement} relies on the fact that, at runtime, we modify only the transition probabilities of $\PIMDP$ that are improved by the new data. Hence, the set of adversaries of the updated \PIMDPText are a subset of those of the \PIMDPText at the previous time step.

Finally, we remark that in the limit as $D$ becomes dense in $X$, 
the regression error in \eqref{eq:GPerror} goes to zero with probability one. 
As a consequence, it is possible to show that as the size of the discretization goes to $0$ and as the coverage of $\dataset$ approaches $X$, Alg. \ref{alg:onlineloop} converges to the optimal strategy for Process \eqref{eq:sys}.

\section{EVALUATIONS}
    \label{sec:evaluation}
    
\begin{figure*}
 \newcommand\subfigwidth{0.6\columnwidth}
\newcommand\figwidth{\columnwidth}
\centering
    \begin{subfigure}{\subfigwidth}
        \centering
        \includegraphics[trim=30 35 0 0,clip,width=\figwidth]{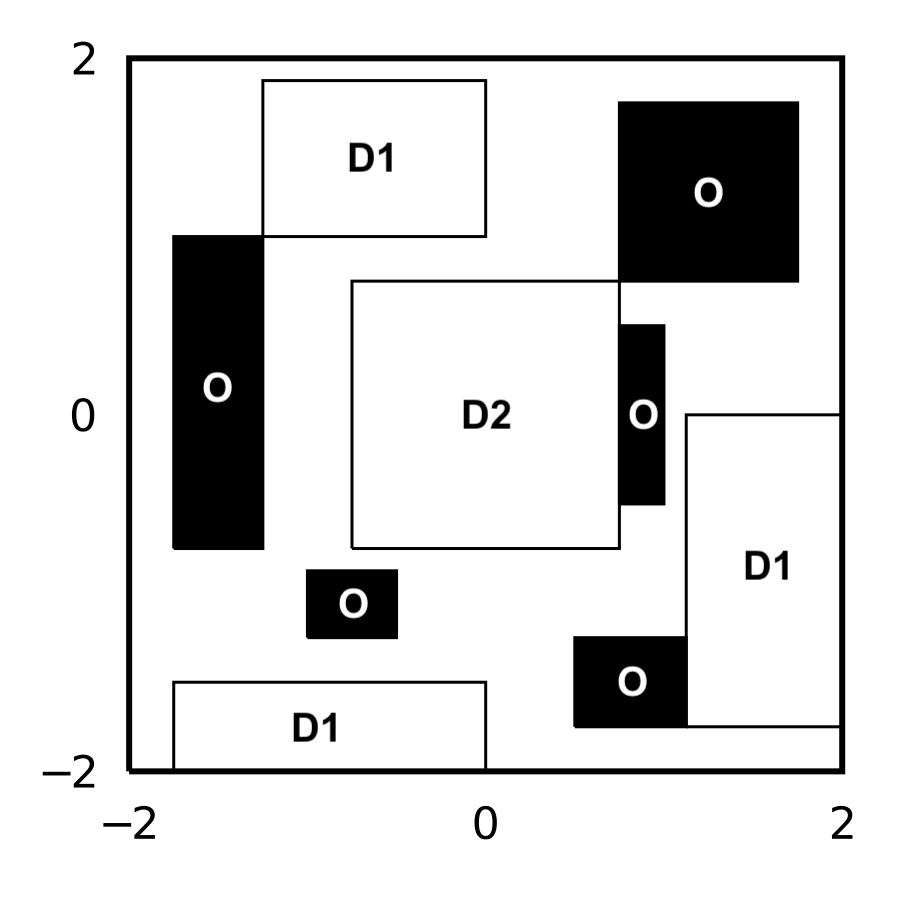}
        \caption{Regions of interests and obstacles}
        \label{fig:ex1}
    \end{subfigure}
    \begin{subfigure}{\subfigwidth}
        \centering
        \includegraphics[trim=30 35 0 0,clip,width=\figwidth]{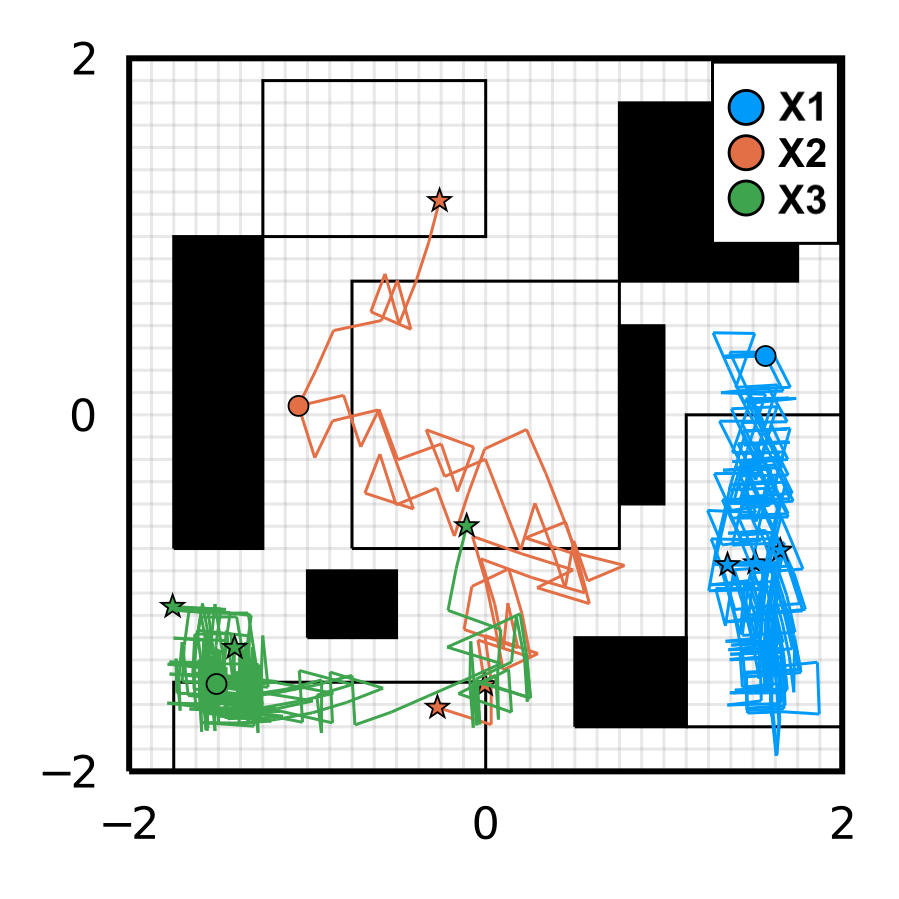}
        \caption{Sink-state metric}
        \label{fig:ex2}
    \end{subfigure}
    \begin{subfigure}{\subfigwidth}
        \centering
        \includegraphics[trim=30 35 0 0,clip,width=\figwidth]{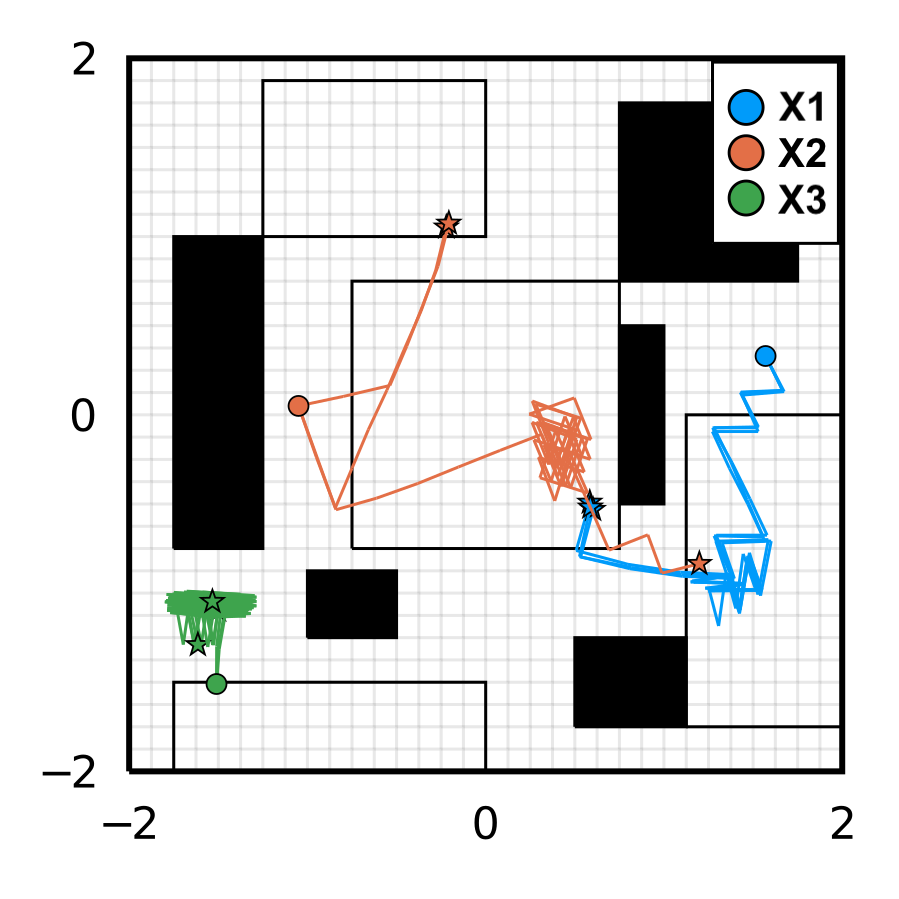}
        \caption{Sink-state and progression metric}
        \label{fig:ex3}
    \end{subfigure}
    \caption{The state space labels and simulations showing the resultant behavior of the two tertiary metrics used for benchmarking for optimal action selection.} \label{fig:examples}
    \vspace{-2mm}
\end{figure*}

\begin{table*}[t]
  \caption{Benchmark results from 500 simulations comparing offline method against the offline-online framework with three types of GP regression methods.  The comparison metrics are: the probabilities of violating and satisfying $\spec$, the average computation time of each control step, and the mean number and value of $\PIMDPText$ transitions updates per run.}
  \label{tab:benchmark-all}
  \centering

\begin{tabular}{cc|ccc|ccccc|ccc}
\hline
\multicolumn{2}{c|}{} & \multicolumn{3}{c|}{Global GP (static)} & \multicolumn{5}{c|}{Local GP (updates with $\nns=75$ 
neighbors)} & \multicolumn{3}{c}{Local GP (static)}\\
\hline
$x_0$ & Tertiary & $P_\text{Violate}$ & $P_\text{Satisfy}$  & Time & $P_\text{Violate}$ & $P_\text{Satisfy}$  & Time & \# $\PIMDP$-updates & Val. $\PIMDP$-updates & $P_\text{Violate}$ & $P_\text{Satisfy}$  & Time \\
\hline
$1$ & Offline & $1.0$ & $0.0$ & $-$ & $1.0$ & $0.0$ & $-$ & $-$ & $-$ & $1.0$ & $0.0$ & $-$\\
$1$ & Sink+Prog & $0.0$ & $1.0$ & $0.0002$ & $0.0$ & $1.0$ & $0.0135$ & 714.8 & 1.0 & $0.0$ & $1.0$ & $0.0003$\\
$1$ & Sink & $0.778$ & $0.216$ & $0.0002$& $\mathbf{0.326}$ & $\mathbf{0.322}$ & $0.0323$ &6258& 0.9996 & $0.77$ & $0.22$ & $0.0004$ \\
\hline
$2$ & Offline & $1.00$ & $0.00$ & $-$ & $1.00$ & $0.00$ & $-$ & $-$ & $-$ & $1.00$ & $0.00$ & $-$ \\
$2$ & Sink+Prog & $0.0$ & $0.76$ & ($0.0001$)  & $0.0$ & $\mathbf{0.996}$ & $0.0222$ &110.8 & 1.0 & $0.0$ & $0.808$ & $0.0001$\\
$2$ & Sink & $0.0$ & $1.0$ & $0.0002$ & $0.0$ & $1.0$ & $0.0115$ &521.6 & 0.9998 & $0.0$ & $1.0$ & $0.0002$\\
\hline
$3$ & Offline & $0.348$ & $0.652$ & $-$ & $0.348$ & $0.652$ & $-$ & $-$ & $-$ & $0.348$ & $0.652$ & $-$\\
$3$ & Sink+Prog & $0.098$ & $0.0$ & $0.0001$  & $0.102$ & $\mathbf{0.864}$ & $0.0130$ &1814& 0.9957& $0.074$ & $0.0$ & $0.0004$\\
$3$ & Sink & $0.0$ & $0.702$ & $0.0001$ & $0.088$ & $0.6140$ & $0.0120$ & 5890 & 0.9970& $0.00$ & $0.704$ & $0.0004$\\
\end{tabular}
    \vspace{-2mm}
\end{table*}

We evaluated our framework on the nonlinear system 
\begin{equation*}
    \x_{k+1} = \x_k + g(\x_k,\u_k) + \procnoise_k
\end{equation*}
where $\u_k \in \{u_1, u_2, u_3, u_4\}$, and $g$ is a priori unknown with $g(\x_k,\u_k) =$
\begin{equation*}
\begin{cases}
[0.25 + 0.05\sin{\x_k^{(2)}}, 0.1\cos{\x_k^{(1)}}]^T & \text{ if } \u_k= u_1\\
[-0.25 + 0.05\sin{\x_k^{(2)}}, 0.1\cos{\x_k^{(1)}}]^T & \text{ if } \u_k = u_2\\
[0.1\cos{\x_k^{(2)}}, 0.25+0.05\sin{\x_k^{(1)}}]^T & \text{ if } \u_k = u_3\\
[0.1\cos{\x_k^{(2)}}, -0.25+0.05\sin{\x_k^{(1)}}]^T & \text{ if } \u_k = u_4,
\end{cases}
\end{equation*}
and $\procnoise_k$ is Gaussian noise drawn from $\mathcal{N}(0, 0.01 I)$.

The compact set $X = [-2,2]^2$ and its regions of interest are shown in Fig.~\ref{fig:ex1}.  The atomic propositions are $O$ (obstacle), $D1$ (destination 1) and $D2$ (destination 2).  The specification is ``to visit destinations 1 and 2 in any order and always avoid the obstacle.''  It translates to the \LTLf formula 
$$\spec=\Globally(\neg O)\wedge \Eventually(D1)\wedge \Eventually(D2).$$

The offline abstraction, \PIMDPText, and $\policyoff$ were computed in an hour with $m=200$ uniformly random datapoints for each action. 
We intentionally used a small $\dataset$ to demonstrate the efficacy of the proposed offline-online control framework.

In the online module, local GPs were created at every step using the $\nns=75$
nearest datapoints and the same hyperparameters as the offline GPs (which were trained using a zero-mean prior and the squared-exponential kernel). 
We chose three initial states from the states with $\plow_{\policyoff} = 0$, and deployed the system with the proposed framework.  
Each simulation ran until the specification was satisfied, violated or the trajectory length reached a predefined bound of 500.  
Note that $\Spec$ is an unbounded-time property, and hence, the trajectories need to be extended to infinite length though this is impossible in practice.
Figure~\ref{fig:examples} shows multiple simulated trajectories generated with the online framework using the sink-state (``Sink'') metric, and the combined sink-state and progression metric (``Sink-Prog'') (see Sec.~\ref{sec:online}). 

Table~\ref{tab:benchmark-all} contains benchmark results for different GP cases and tertiary metrics. We compare the empirical probabilities of violating and satisfying $\spec$ using \emph{static global GPs}, \emph{static local GPs}, and \emph{local GPs with online updates} over 500 simulations.
The framework is evaluated using the tertiary metrics for each GP case and compared to the performance of $\policyoff$, which has $\plow_{\policyoff}(s) = 0$ for all possible initial states.
In other words, $\policyoff$ does not provide any \textit{a priori} satisfaction guarantees.

First, we compare the efficacy of our online framework and the different tertiary metrics against the offline strategy.
In all cases, the online framework decreases the number of $\spec$-violating runs, and in many cases increases the number of $\spec$-satisfying runs.  
Notably for the third 
initial state, $\policyoff$ results in a 35-65 split in violating and satisfying runs while the online framework sees consistent drops in the percentage of violating runs.
When the sum of the fractions of violating and satisfying runs is less than one, the remainder runs terminated in a Zeno-like behavior as observed in Figure~\ref{fig:ex2}.  
This may be a beneficial behavior as it leads to collection of more data and reduction in the regression error.

The three major columns of Table~\ref{tab:benchmark-all} compare the outcomes using three types of GPs.
In many cases, particularly for the second 
initial state, \emph{local GPs with updates}  provides a benefit to the number of $\spec$-satisfying runs. 
The proportion of $\spec$-violating runs is worse for the third 
initial state, but the most successful runs are realized using the local GPs and the combined tertiary metrics.
The same benchmark was run on the global GP with online updates (not included in the table), but all the trials timed out after 24 hours with no result clearly illustrating the advantage of local GPs for online computations.

Constructing the local GPs without updating them or the \PIMDPText is only marginally slower than using the global GP directly.
The additional time to update the local GP and update the \PIMDPText transitions is significant compared to the static methods, but are still reasonable for online scenarios. 
Collecting data online and updating the \PIMDPText abstraction takes time, but it results in improved transition intervals between states. 
The additional data collected while using the sink-state metric shows many intervals are improved due to long trajectories that neither violate nor satisfy $\Spec.$

These evaluations exhibit the advantage of using the proposed offline-online control framework.  
The framework was able to increase the probability of satisfaction, in some cases from 0\% to 100\%, and the use of local GPs enables us to overcome computational limitations.
Further investigation is needed to address the aforementioned hyperparameter choices, e.g., the value of $\nns$, and number of $\PIMDP$ states to update.

\section{CONCLUSION}
    \label{sec:conclusion}
In this work, we presented a synergistic offline-online control framework for stochastic systems with an unknown component via local GP regression.  The offline module provides a baseline controller and guarantees for the online module.  Online, the controller and its guarantees are iteratively refined as more data is collected.  
Evaluations illustrate the advantage of the framework over just the offline method.  Future directions include investigation into hyperparatmeter choices for the online method and experiments (deployments) of actual physical platforms with this framework.

\bibliographystyle{ieeetr}
\bibliography{references,cite,lahijanian}


\end{document}